\documentclass[%
reprint,
showkeys,
nofootinbib,
amsmath,amssymb,
aps,
prd,
]{revtex4-2}

\usepackage{graphicx}
\usepackage{subcaption}
\usepackage[colorlinks,allcolors=black,CJKbookmarks=True]{hyperref}

\renewcommand{\d}{{\rm d}}

\begin{document}

\title{Loss of the scaling attractor in self-gravitating domain wall networks}

\author{Zhen-Min Zeng}
\email{zzhenmin@itmp.msu.ru}
\affiliation{
Institute for Theoretical and Mathematical Physics, MSU, 119991 Moscow, Russia}

\begin{abstract}

Domain-wall(DW) networks are known to approach a relativistic scaling regime on fixed radiation- and matter-dominated backgrounds, forming the basis of the no-frustration conjecture. However, this picture assumes that the defect network remains gravitationally subdominant. We investigate the self-consistent evolution of DWs by coupling the velocity-dependent one-scale model to the Friedmann equation and radiation energy transfer. The resulting autonomous system allows the cosmic expansion history to evolve dynamically rather than being imposed externally. We demonstrate analytically that gravitational backreaction qualitatively changes the phase-space structure: the radiation-era scaling solution, which is a stable attractor on a fixed background, becomes a saddle once the expansion rate is promoted to a dynamical degree of freedom. Furthermore, we establish that no stable fixed point exists within the physical phase space. Consequently, the scaling regime survives only as a transient stage, and all trajectories are driven toward a wall dominated and kinematically frustrated state in which the walls freeze in comoving coordinates. Our results demonstrate that the scaling attractor is not preserved in self-gravitating DW networks and reveal the generic late-time frustration dynamics of wall domination.

\end{abstract}
\maketitle

\section{Introduction}
Domain walls(DWs) are two-dimensional topological defects formed during spontaneous breaking of a discrete symmetry~\cite{Kibble:1976sj,Vilenkin:1984ib}. Their formation is a generic prediction of many scenarios beyond-the-Standard-Model, including axion models~\cite{Sikivie:1982qv,Linde:1990yj,Vilenkin:2000jqa,Hiramatsu_2011,Kawasaki:2014sqa,Hiramatsu:2012gg,Hiramatsu_2013}, grand unified theories~\cite{Bhattacharjee:1991zm,Bajc:1997ky,Chakrabortty:2019fov,King_2021}, and various mechanisms of dynamical symmetry breaking~\cite{Abel:1995wk,Preskill:1991kd,Battye:2011jj}. Historically, stable DWs have been regarded as a cosmological catastrophe~\cite{Zeldovich:1974uw}, since they will eventually dominate the Universe, which is strictly constrained by observation~\cite{ferreira2025axionicdefectscmbbirefringence}. To avoid this outcome, any realistic model must ensure that the network eventually decays, for example, through vacuum bias~\cite{Gelmini:1988sf,PhysRevD.53.4237,PhysRevD.28.1419,Babichev:2025stm,Wei:2022poh}, biased initial conditions~\cite{Lalak:1993ay,Gonzalez:2022mcx}, inverse phase transition~\cite{Lew_1993,Dvali_1995,Babichev:2021uvl,Li:2023tdx}, or quantum tunneling~\cite{Blasi_2022}. The decay of domain-wall networks can generate a stochastic gravitational-wave background with characteristic spectral features, potentially observable by current and future detectors~\cite{PhysRevD.23.852,Chang_1998,Gleiser_1998,Hiramatsu_2010,Kuroyanagi_2018,Dankovsky:2024ipq,Dankovsky:2024zvs,Saikawa_2017,roshan2026imprintdomainwallannihilation,Notari:2025kqq,blasi2025domainwallsscalingregime,cyr2025nearpeakspectrumgravitationalwaves}.

Although DWs must eventually decay, they may nevertheless play an important role in the thermal history of the Universe. At present, there is no observational constraint excluding a phase in which DWs temporarily dominate the cosmic energy density, provided that they decay before the onset of Big Bang Nucleosynthesis. Such a scenario can naturally arise for domain walls formed at sufficiently high energy scales. It is therefore important to understand their full dynamical evolution history. 

Large-scale lattice simulations~\cite{1989ApJ...347..590P,Avelino_2006,PhysRevD.84.103523,Garagounis_2003,Oliveira_2005,Leite_2013,avelino2022analyticalscalingsolutionsevolution} have shown, and the velocity-dependent one-scale (VOS) model~\cite{Avelino_2005,Martins_2016} has successfully reproduced, that domain-wall networks evolve toward a relativistic scaling regime on fixed cosmological backgrounds, characterized by a characteristic length scale proportional to cosmic time ($L\propto t$) and an approximately constant root-mean-square velocity(RMS) ($v\rightarrow \mathrm{const}$). This behavior underlies the no-frustration conjecture~\cite{Avelino_2006}, according to which realistic domain-wall networks do not freeze in comoving coordinates.

Existing studies of the no-frustration conjecture implicitly assume that the defect network remains gravitationally subdominant, with the expansion history imposed externally through $a\propto t^\lambda$. Under this fixed-background assumption, the scaling solution is indeed a stable attractor~\cite{Avelino_2006,avelino2022analyticalscalingsolutionsevolution}. However, the fixed-background assumption cannot be maintained at all times, because the wall energy density dilutes as $\rho_w\propto t^{-1}$, more slowly than radiation ($\rho_r\propto t^{-2}$). Consequently, the wall network starts dominating cosmic expansion, even if it is subdominant initially. Although freezing-like behavior after wall domination is often qualitatively anticipated, the dynamical fate of the scaling solution itself -- whether it survives as an attractor once the expansion rate becomes a self-consistent variable---has not previously been analyzed within a closed autonomous system.

The central question is therefore not whether scaling
exists on a radiation background—it does—but whether
it remains the late-time attractor once they start to dominate the Universe.

In this work, we show that including gravitational backreaction qualitatively changes the phase-space structure. Coupling the VOS equations with the Friedmann equation with energy transfer from walls to radiation produces a closed autonomous dynamical system in which the expansion-rate variable $q\equiv Ht$ becomes dynamical. We find that the radiation-era scaling fixed point---a genuine attractor on a fixed background---becomes a saddle due to the coupled expansion-rate degree of freedom. A stability analysis further shows that no stable fixed point exists in the physical phase space when radiation constitutes only a fraction of the energy budget. The scaling regime therefore cannot be the asymptotic fate of the system: it survives only as a transient, saddle-controlled epoch.

The physical mechanism originates from the backreaction of DWs on the background. As the wall fraction grows, the effective equation of state softens from $w_{\rm eff}\approx 1/3$ to $-2/3$, the deceleration weakens, and the increased Hubble friction suppresses the RMS velocity. Lower velocities reduce energy injection into radiation, further accelerating the decline of the radiation fraction. This feedback drives the system toward a runaway asymptotic branch with $v\to0$, $L\propto a$, and $q\to2$, corresponding to kinematic frustration. In essence, scaling dynamically destroys the radiation-dominated background that originally sustains it.

The paper is organized as follows. Section~\ref{sec:vos-fixed} reviews the VOS model on a fixed radiation background and confirms the scaling attractor, recovering the no-frustration result as a baseline. Section~\ref{sec:self-grav} constructs the self-consistent VOS-Friedmann system and presents the stability analysis. Section~\ref{sec:numerical} presents numerical integration and the global phase portrait. Section~\ref{sec:conclusions} discusses implications, limitations, and connections to frozen-network cosmology.

\section{Standard VOS Model on a Fixed Radiation Background}
\label{sec:vos-fixed}
The VOS model is the standard effective description of defect network evolution and has been extensively calibrated against large-scale simulations. Although it is not a systematic expansion with a controlled small parameter, it provides an accurate macroscopic description over a broad range of fixed cosmological backgrounds.

We begin with the standard VOS analysis, in which the universe's expansion is prescribed externally and the defect network is treated as a spectator field. The purpose of this section is to establish the VOS equations, define notation, and demonstrate that on a fixed radiation background the scaling solution is a genuine attractor --- thereby recovering the no-frustration result within the VOS framework. This serves as the baseline for comparison when backreaction is introduced in Sec.~\ref{sec:self-grav}.

The VOS model provides a coarse-grained statistical description of a defect network in terms of two variables: the characteristic length scale $L$ (or equivalently the energy density $\rho_w = \sigma_w/L$, where $\sigma_w$ is the wall tension) and the RMS $v$. The model can be derived by averaging the Nambu-Goto equations of motion over a statistical ensemble of wall segments; we state the final equations and refer to~\cite{Avelino_2005,Martins_2016} for the full microscopic derivation.

The energy density of a wall network evolves under the competing effects of cosmological expansion, curvature-driven acceleration, and energy loss through wall intersections, which can be represented as
\begin{align}
\frac{\d L}{\d t} &= HL(1 + 3v^{2}) + c_w v, \label{eq:vos-L} \\
\frac{\d v}{\d t} &= (1 - v^{2})\left(\frac{k_w}{L} - 3Hv\right). \label{eq:vos-v}
\end{align}
Here $L$ characterizes the correlation length scale (or mean separation distance) of the wall network. $v$ is the RMS velocity of the DWs.
$H$ is the Hubble parameter. $c_w$ and $k_w$ represent the efficiency of energy loss and effect curvature, which we assume to be constant.  

The parameters $c_w$ and $k_w$ can be calibrated against field-theory lattice simulations; typical values in the radiation era are $c_w \sim 0.66$ and $k_w \sim 0.81$~\cite{Martins_2016}. For the radiation-dominated regime considered in this section, the constant-parameter approximation is well supported by simulations.

Eq.~(\ref{eq:vos-L}) combines the dilution due to Hubble expansion $HL$ and the stretching of relativistic walls $3H L v^2$ with a phenomenological energy-loss term $c_w v$. Eq.~(\ref{eq:vos-v}) encodes the competition between acceleration from wall curvature $k_w/L$ and Hubble friction $3Hv$.

On a fixed FRW background, $H = \lambda / t$ with $\lambda = 1/2$ (radiation) or $\lambda = 2/3$ (matter).\footnote{In this section $\lambda$ denotes the fixed expansion index $a \propto t^\lambda$. In Sec.~\ref{sec:self-grav} we use $q \equiv Ht$ as a dynamical variable.} Introducing $x \equiv L/t$ and $N \equiv \ln t$ (with $' \equiv \d/\d N$), the VOS equations become:
\begin{align}
x' &= x[\lambda(1+3v^2) - 1] + c_w v, \label{eq:fx-x} \\
v' &= (1-v^2)\left(\frac{k_w}{x} - 3\lambda v\right). \label{eq:fx-v}
\end{align}

A scaling solution corresponds to a fixed point $(x_*, v_*)$. Applying $v' = 0$ and $x' = 0$, we have:
\begin{equation}
v_*^2 = \frac{(1-\lambda)k_w}{3\lambda(k_w + c_w)}, \qquad
x_* = \frac{2k_w}{3v_*}.
\label{eq:rd-fixed-vos}
\end{equation}

Linearizing around the fixed point with $x = x_* + \delta x$, $v = v_* + \delta v$, the Jacobian is:
\begin{equation}
M = \begin{pmatrix}
\lambda(1+3v_*^2)-1 & 6\lambda x_* v_* + c_w \\[4pt]
-(1-v_*^2)\dfrac{k_w}{x_*^2} & -3\lambda(1-v_*^2)
\end{pmatrix}_{(x_*,v_*)}.
\label{eq:jacobian-standard}
\end{equation}
Using Eq.~(\ref{eq:rd-fixed-vos}), it is easy to see that  $\operatorname{Tr}(M) < 0$ and $\det(M)  > 0$ if $0<\lambda<1$ and $c_w,k_w>0$. A $2 \times 2$ real matrix with $\operatorname{Tr} < 0$ and $\det > 0$ has eigenvalues satisfying $\Re(\mu_{\pm}) < 0$. The fixed point is therefore a stable attractor: perturbations decay as $\delta X \propto e^{\mu N}$ with $\Re(\mu) < 0$.

Physically, the stability reflects a balance between curvature acceleration ($k_w/L$), Hubble friction ($3Hv$), and energy loss ($c_w v/L$). A deviation from scaling triggers restoring dynamics that push the network back, explaining why DW networks approach scaling on a fixed radiation background --- consistent with the no-frustration conjecture. Crucially, this entire analysis assumes $H(t)$ is externally prescribed; the walls never contribute significantly to the total energy density.

\section{Self-Consistent VOS-Friedmann System}
\label{sec:self-grav}

The analysis of Sec.~\ref{sec:vos-fixed} assumes the defect energy density is gravitationally subdominant. For DWs, $\rho_w = \sigma/L$, and in scaling $\rho_w \propto t^{-1}$ while radiation dilutes as $\rho_r \propto t^{-2}$. The ratio $\Omega_w/\Omega_r \propto t$ grows without bound. Once $\Omega_w$ becomes non-negligible, the fixed-background framework fails: $H$ is determined by the wall density through $H^2 = (8\pi G/3)(\rho_w + \rho_r)$; the energy lost by walls must be injected into a radiation bath, closing the feedback loop; and $L \propto t$ is no longer guaranteed. We retain the FRW metric as a large-scale coarse-grained effective geometry, without claiming local isotropy where anisotropic wall stresses are important.

From $\rho_w = \sigma/L$ and the VOS Eq.~(\ref{eq:vos-L}):
\begin{equation}
\dot\rho_w = -\rho_w\frac{\dot L}{L} = -\rho_w\Bigl[H(1+3v^2) + \frac{c_w v}{L}\Bigr].
\end{equation}
We define the effective equation-of-state parameter $\omega_w$ through the continuity equation.
Comparing with the continuity equation $\dot\rho_w + 3H(1+\omega_w)\rho_w = -(c_w v/L)\rho_w$, we read off the effective EOS:
\begin{equation}
\omega_w = v^2 - \frac{2}{3}.
\label{eq:eos-wall}
\end{equation}
In the relativistic limit ($v \to 1$) the network behaves as a relativistic gas ($\omega_w \to 1/3$); in the frozen limit ($v \to 0$) it dilutes as $\rho_w \propto a^{-1}$ ($\omega_w = -2/3$).

The energy lost by walls is injected into a radiation bath. In the homogeneous effective fluid approximation, the two-component continuity equations are:
\begin{align}
\dot\rho_w + 3H(1+\omega_w)\rho_w &= -\frac{c_w v}{L}\rho_w, \label{eq:cont-w} \\
\dot\rho_r + 4H\rho_r &= +\frac{c_w v}{L}\rho_w. \label{eq:cont-r}
\end{align}
The system is closed by the Friedmann equation $H^2 = (8\pi G/3)(\rho_w + \rho_r)$.

To avoid presupposing $L \propto t$, we adopt variables that make no assumption about the scaling law:
\begin{equation}
y \equiv HL, \qquad q \equiv Ht, \qquad \Omega_r \equiv \frac{\rho_r}{\rho_w + \rho_r},
\label{eq:vars-new}
\end{equation}
with $\rho_w = \sigma/L$. The physical phase space is bounded by $y > 0$, $0 \le v \le 1$, $q > 0$, $0 \le \Omega_r \le 1$. In radiation dominated(RD) scaling, $y = \text{const}$ and $q = 1/2$; in a frozen-wall regime, $y \propto t$ and $q \to 2$.

Substituting $\omega_w = v^2 - 2/3$ and using the Friedmann equation, the closed autonomous system is (see Appendix~\ref{app:derivation}):
\begin{equation}
\begin{aligned}
y' &= yq\Bigl[\frac{1}{2} + \frac{3}{2}v^2(1+\Omega_r) - \frac{3}{2}\Omega_r\Bigr] + c_w v q, \\[1ex]
v' &= (1-v^2)q\Bigl(\frac{k_w}{y} - 3v\Bigr), \\[1ex]
q' &= q - \frac{3}{2}\Bigl[\frac{1}{3} + \Omega_r + v^2(1-\Omega_r)\Bigr]q^2, \\[1ex]
\Omega_r' &= q(1-\Omega_r)\Bigl[\frac{c_w v}{y} - 3\Omega_r(1-v^2)\Bigr].
\end{aligned}
\label{eq:4d-system}
\end{equation}

We determine all fixed points of~(\ref{eq:4d-system}) within the physical domain $y > 0$, $0 < v < 1$, $q > 0$, $0 \le \Omega_r \le 1$, with $c_w > 0$, $k_w > 0$.

Since Eq.~(\ref{eq:4d-system}) is a 4D system, to analyze its fix point($X_i'=0$, with $X = (y, v, q, \Omega_r)$) and linear instability, we need to evaluate the $4 \times 4$ Jacobian matrix $J_{ij} = \partial X_i'/\partial X_j$  at the fixed point and computing the characteristic polynomial $\det(J - \mu I) = \mu^4 + a_1 \mu^3 + a_2 \mu^2 + a_3 \mu + a_4$. The Routh-Hurwitz criterion for asymptotic stability requires:
\begin{equation}
a_1 > 0,\quad a_2 > 0,\quad a_3 > 0,\quad a_4 > 0,
\label{eq:hurwitz-basic}
\end{equation}
\begin{equation}
H_1 \equiv a_1 a_2 - a_3 > 0,
\label{eq:hurwitz-H1}
\end{equation}
\begin{equation}
H_2 \equiv a_1 a_2 a_3 - a_1^2 a_4 - a_3^2 > 0.
\label{eq:hurwitz-H2}
\end{equation}

If all the inequalities above are satisfied, then the four eigenvalues of the Jacobian matrix will all have negative real parts, which implies that the fixed point is stable.

A fixed point requires $\Omega_r' = 0$, which can be satisfied on the boundaries $\Omega_r = 0$, $\Omega_r = 1$, or at an interior point.

\textit{Radiation boundary ($\Omega_r = 1$).} Equations~(\ref{eq:4d-system}) reduce to the fixed-background VOS system. A unique finite fixed point exists:
\begin{equation}
q_* = \frac{1}{2}, \qquad
v_*^2 = \frac{k_w}{3(k_w + c_w)}, \qquad
y_* = \frac{k_w}{3v_*}.
\label{eq:fp-rd-4d}
\end{equation}
Linear stability analysis (Appendix~\ref{app:rd-stability}) reveals that this point is a saddle in the full 4D system. The Jacobian possesses a single unstable eigenvalue,
\begin{equation}
\mu = +1,
\label{eq:mu-plus-one}
\end{equation}
which is independent of the VOS parameters $c_w$ and $k_w$. While perturbations in the $(y, v)$ subspace decay (recovering the fixed-background attractor of Sec.~\ref{sec:vos-fixed}), the $\mu=+1$ direction --- originating from the self-coupling of the expansion-rate variable $q$ --- renders the point unstable. The parameter independence of this eigenvalue indicates that the instability is not a calibration artifact: it is a structural consequence of promoting the expansion rate from an externally prescribed function to a dynamical variable.

\textit{Wall boundary ($\Omega_r = 0$).} At $\Omega_r = 0$, $\Omega_r' = c_w v q / y$, which vanishes only if $v = 0$. But then $v' = 0$ requires $y = k_w/3v \to \infty$. No finite fixed point exists on this boundary.

\textit{Interior ($0 < \Omega_r < 1$).} Setting $\Omega_r' = 0$ with $0<\Omega_r<1$ (so $1-\Omega_r\neq0$) and $q>0$ gives:
\begin{equation}
\frac{c_w v}{y} - 3\Omega_r(1-v^2) = 0,
\end{equation}
which, combined with $y = k_w/(3v)$ from $v' = 0$, yields:
\begin{equation}
\Omega_r = \frac{c_w v^2}{k_w(1-v^2)}.
\label{eq:omegar-interior}
\end{equation}
The $q' = 0$ condition then gives:
\begin{equation}
q_* = \frac{2}{1 + 3v^2\left(1 + \frac{c_w}{k_w}\right)}.
\label{eq:q-interior}
\end{equation}

We now proceed by contradiction. Assuming that an interior fixed point is asymptotically stable and imposing the basic positivity conditions~(\ref{eq:hurwitz-basic}) together with $0 < v < 1$, $c_w > 0$, $k_w > 0$, and then adding $H_1 > 0$ and $H_2 > 0$ sequentially, the system reduces to a compact necessary condition for stability:
\begin{equation}
k_w < \frac{v^2 c_w}{1 - v^2},
\qquad 0 < v < \sqrt{\frac{2}{3}}.
\label{eq:kw-stability-bound}
\end{equation}
(See Appendix~\ref{app:jacobian-interior} for the algebraic derivation.)

Combining the stability bound~(\ref{eq:kw-stability-bound}) with the fixed-point condition~(\ref{eq:omegar-interior}):
\begin{equation}
\Omega_r = \frac{c_w v^2}{k_w(1-v^2)} > 1.
\label{eq:omegar-gt-1}
\end{equation}
Thus, any parameter combination satisfying the necessary Hurwitz stability conditions automatically yields $\Omega_r > 1$. Since the physical phase space is bounded by $0 \le \Omega_r \le 1$, the mathematically stable interior fixed point lies strictly outside the physically admissible region.

Thus, within the physical phase space $0 \le \Omega_r \le 1$, no stable fixed point exists anywhere: the interior point is forced to $\Omega_r > 1$ for stability, and the RD boundary point is a saddle.

\textit{Asymptotic runaway.} With no stable attractor in the physical phase space, any trajectory eventually departs from both the RD boundary and the interior region. The $\Omega_r'$ equation,
\begin{equation}
\Omega_r' = q(1-\Omega_r)\Bigl[\frac{c_w v}{y} - 3\Omega_r(1-v^2)\Bigr],
\end{equation}
reveals two competing effects: energy injection from decaying walls ($c_w v/y>0$) drives $\Omega_r$ upward, while Hubble dilution $-3\Omega_r(1-v^2)<0$ (for $v<1$) drives it downward. When $\Omega_r$ is not extremely small, the dilution term dominates and $\Omega_r'<0$. The system flows toward $\Omega_r \to 0$, $v \to 0$, $y \to \infty$ --- the runaway branch.

In this limit ($\Omega_r \ll 1$, $v \ll 1$), the leading-order equations are:
\begin{align}
v' &\simeq -3q v, \qquad\;\,\Longrightarrow\; v \to 0, \\
y' &\simeq \tfrac{1}{2}yq, \qquad\quad\; \Longrightarrow\; y \propto t, \\
q' &\simeq q\Bigl(1 - \frac{q}{2}\Bigr),
\end{align}
so $q \to 2$ asymptotically, consistent with $\omega_w \to -2/3$ ($a \propto t^2$). From $L = y/qH$ with $H = q/t$ and $q \to 2$, $y \propto t$, we obtain $L \propto t^2 \propto a$, corresponding to kinematic frustration: the network freezes in comoving coordinates with $v \to 0$ and $L/a = \text{const}$.

In summary: the RD fixed point is a saddle, and the interior fixed point can be stable only for $\Omega_r > 1$. No stable attractor exists in the physical phase space. The above analysis assumes constant $c_w, k_w$, but the proof is structurally robust: if a scaling attractor existed in $0<\Omega_r<1$, the phenomenological parameters --- whatever their microphysical dependence on $v$ or the background --- would asymptote to constant values at that fixed point, reducing to the case analyzed here. The $\mu=+1$ eigenvalue at the RD boundary is parameter-independent. The instability is therefore not an artifact of the constant-parameter approximation but a robust consequence of the autonomous structure. Numerical integration (Sec.~\ref{sec:numerical}) confirms this picture.

\section{Numerical Results}
\label{sec:numerical}

We integrate the system Eqs.~(\ref{eq:4d-system}) numerically from $N = 0$ to $N = 15$ ($N\equiv \ln t$)with VOS parameters $c_w = 0.81$, $k_w = 0.66$, calibrated by Martins et al.~\cite{Martins_2016} from $4096^3$ lattice simulations. Here, time $t$ is normalized by its initial value $t_i$, yielding the dimensionless time $\tilde{t} = t/t_i$. Initial conditions correspond to deep radiation domination: $\Omega_r = 0.999$, with $v$, $y$, $q$ at the RD fixed-point values~(\ref{eq:fp-rd-4d}), namely $v_i = v_* = 0.387$, $y_i = y_* = 0.569$, $q_i = 1/2$.

The analytical study of Sec.~\ref{sec:self-grav} established that the RD point is a saddle and that there is no stable fixed point in the physical phase space. We therefore begin with the global phase portrait, which directly confirms that the runaway is generic rather than an artifact of a particular initial condition.

\begin{figure}[h]
    \centering
    \includegraphics[width=\columnwidth]{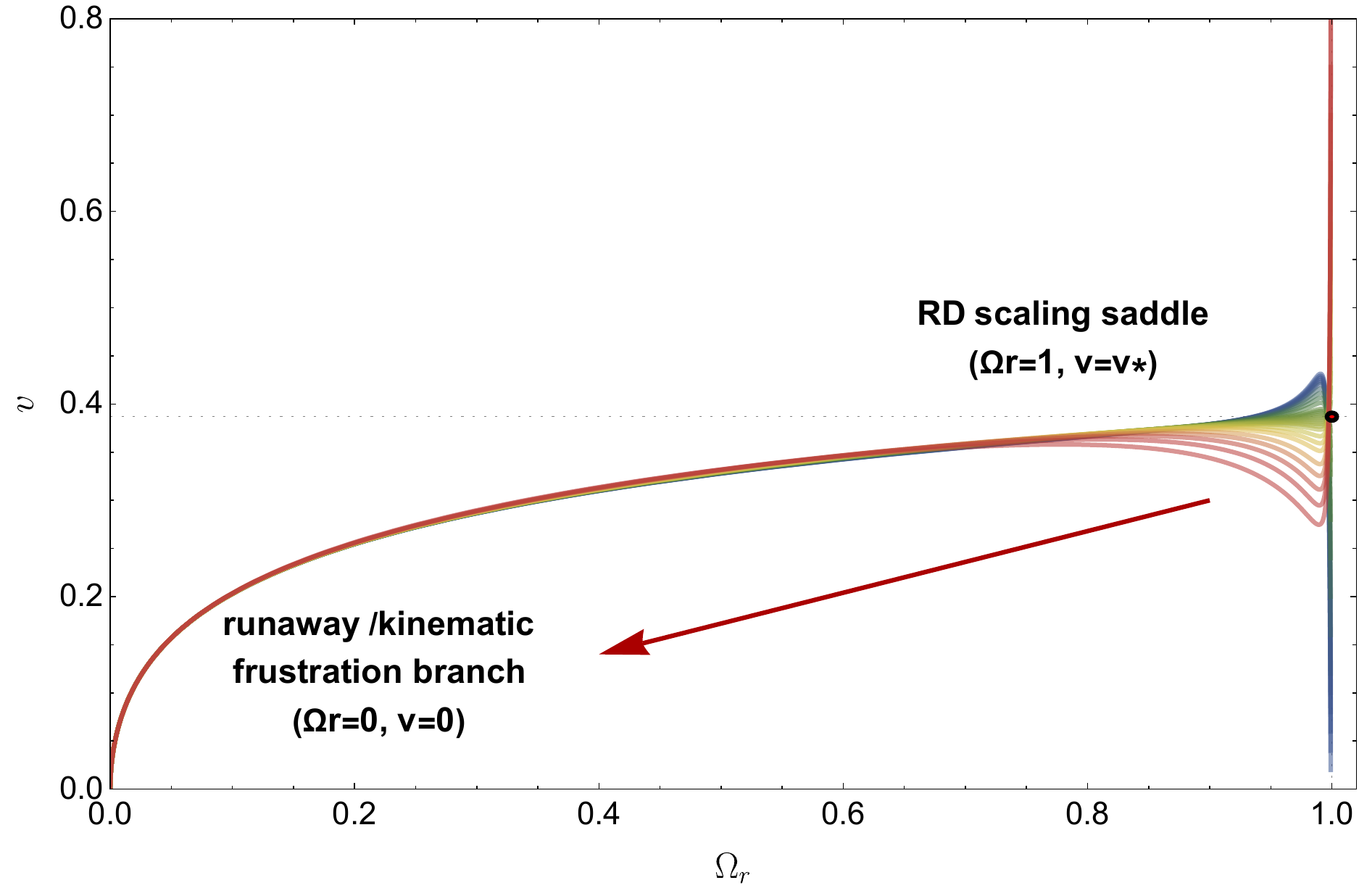}
    \caption{Phase portrait in the $(\Omega_r, v)$ plane. The red disk marks the RD scaling saddle at $(\Omega_r=1, v=v_*)$. All trajectories are repelled along the unstable direction and flow toward $(\Omega_r, v)\to(0,0)$, corresponding to kinematic frustration.}
    \label{fig:phase_portrait}
\end{figure}

The phase portrait (Fig.~\ref{fig:phase_portrait}) shows 28 trajectories with identical physical initial conditions ($\Omega_{r,i}=0.999$, $y_i=y_*$, $q_i=1/2$) but different initial RMS velocities $v_i \in [0.02, 0.85]$. It illustrates the dynamical role of the RD point geometrically transparent. Trajectories with $v_i$ near $v_*$ linger near the saddle for several e-folds before being repelled; those farther from $v_*$ depart more promptly. Crucially, every trajectory ultimately flows toward $(\Omega_r, v) \to (0,0)$. This is the numerical counterpart of the analytic result of Sec.~\ref{sec:self-grav}: the eigenvalue $\mu = +1$ guarantees the RD point is a saddle, and the absence of a stable attractor in the physical phase space forces all trajectories toward the runaway branch.

To examine the detailed evolution, we select a representative trajectory with the canonical RD initial condition ($v_i = v_*$, $y_i = y_*$, $q_i = 1/2$, $\Omega_{r,i}=0.999$).

\begin{figure}
    \centering
    \includegraphics[width=\columnwidth]{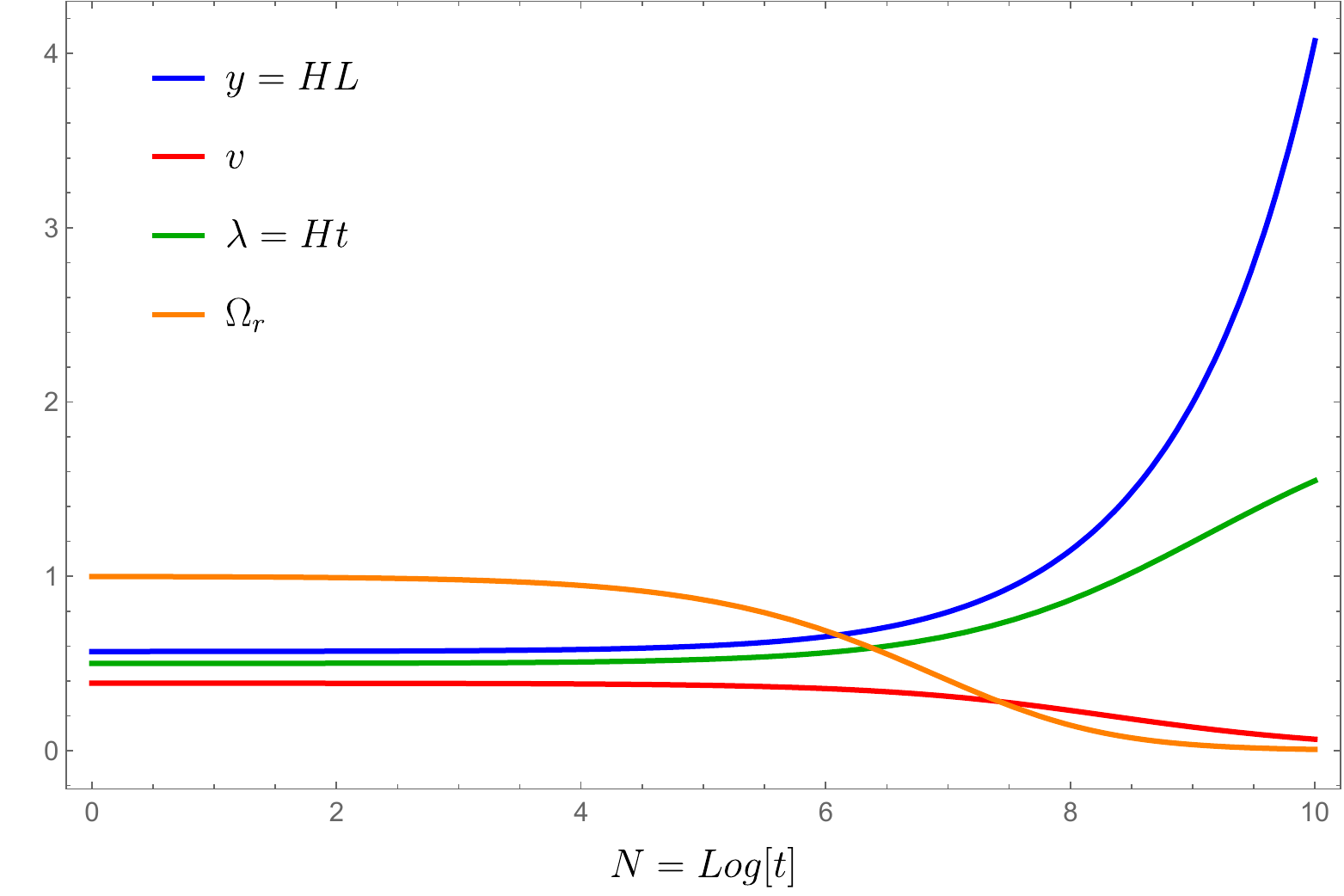}
    \caption{Evolution of the autonomous variables along the representative trajectory. During the RD epoch the system tracks the saddle; as $\Omega_r$ drops, $v$ is damped toward zero, $y$ grows, and $q$ evolves from $1/2$ to $2$, consistent with the transition from radiation-like to wall-dominated expansion.}
   \label{fig:autonomous}
\end{figure}

Fig.~\ref{fig:autonomous} shows the evolution of the autonomous variables for this trajectory. During the RD epoch ($N \lesssim 6$), $\Omega_r \approx 1$ and $(y, v)$ remain near their fixed-point values. Once $\Omega_r$ drops below $\sim 0.5$ at $N \sim 7$--$8$, the system departs from the saddle: $v$ decays to $\sim 5.7 \times 10^{-4}$, $y$ grows to $\sim 466$, and $q$ transitions from $1/2$ toward $2$, consistent with the expansion law evolving from radiation-like to wall-dominated.

The wall fraction $\Omega_w = 1 - \Omega_r$ grows as $\Omega_w \propto t$ during RD, with equality $\Omega_w = \Omega_r$ reached at $N \sim 8$ and $\Omega_w > 0.99$ by $N \sim 10$. After the transition, $\rho_w$ overtakes $\rho_r$ and dilutes as $\rho_w \propto a^{-1} \propto t^{-2}$, consistent with $a \propto t^2$.

\begin{figure*}
    \centering
    \begin{subfigure}[b]{0.45\textwidth}
       \includegraphics[width=\textwidth]{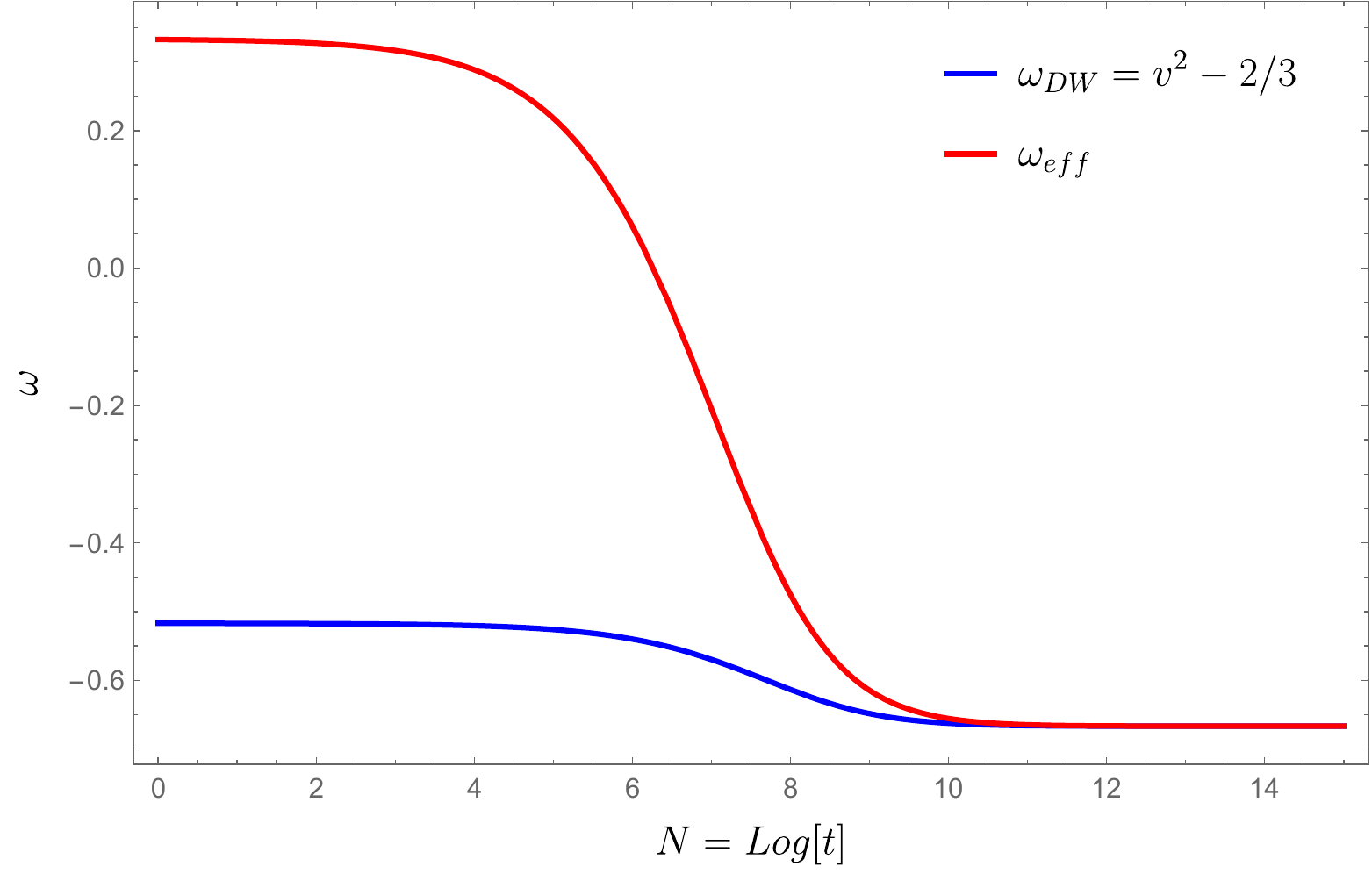}
       \caption{}
       \label{fig:omegaeff}
    \end{subfigure}
    \hfill
    \begin{subfigure}[b]{0.45\textwidth}
        \includegraphics[width=\textwidth]{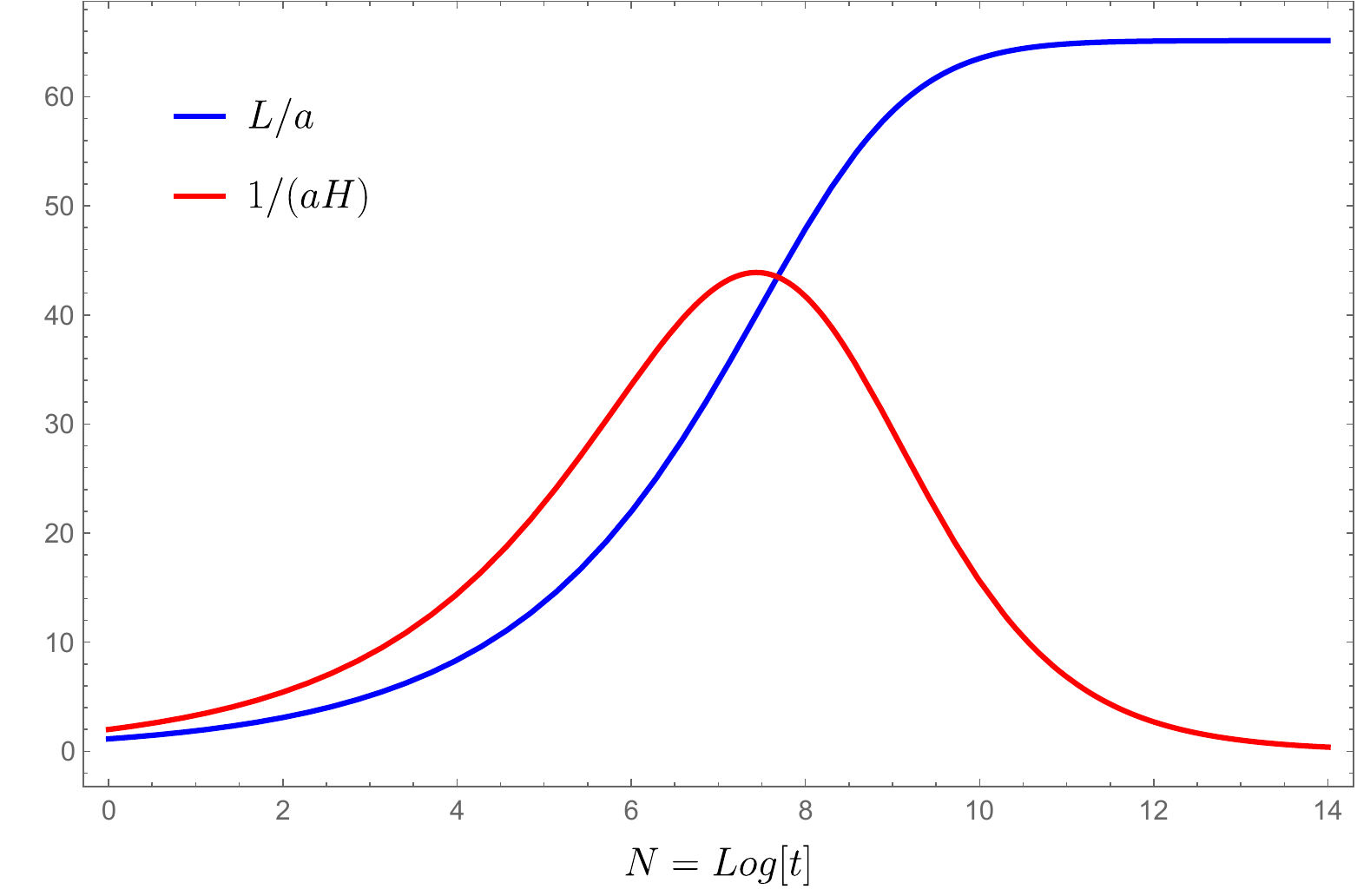}
        \caption{}
        \label{fig:scale}
    \end{subfigure}
    \caption{Left: Evolution of $w_{\rm DW}$ and $w_{\rm eff}$, which both approaching $-2/3$ asymptotically. Right: Evolution of comoving correlation length $L$ and comoving Hubble radius $H^{-1}$. In the asymptotic regime $L \propto a$, confirming kinematic frustration.}
    \label{fig:asymptotic}
\end{figure*}

The asymptotic state is verified in Fig.~\ref{fig:asymptotic}. At $N = 0$, $w_{\rm eff} \approx 1/3$ (radiation), while at $N = 15$, $w_{\rm eff} \approx -2/3$, matching a frozen-wall network with $v \to 0$ and $w_{\rm DW} = v^2 - 2/3 \to -2/3$. The right panel compares $L$, $a$, and $H^{-1}$: during RD, $L/H^{-1} = y \approx \text{const}$, but after the transition $L$ decouples from the Hubble scale and grows as $L \propto a$, the defining signature of kinematic frustration.

\begin{figure}
    \centering
    \includegraphics[width=\columnwidth]{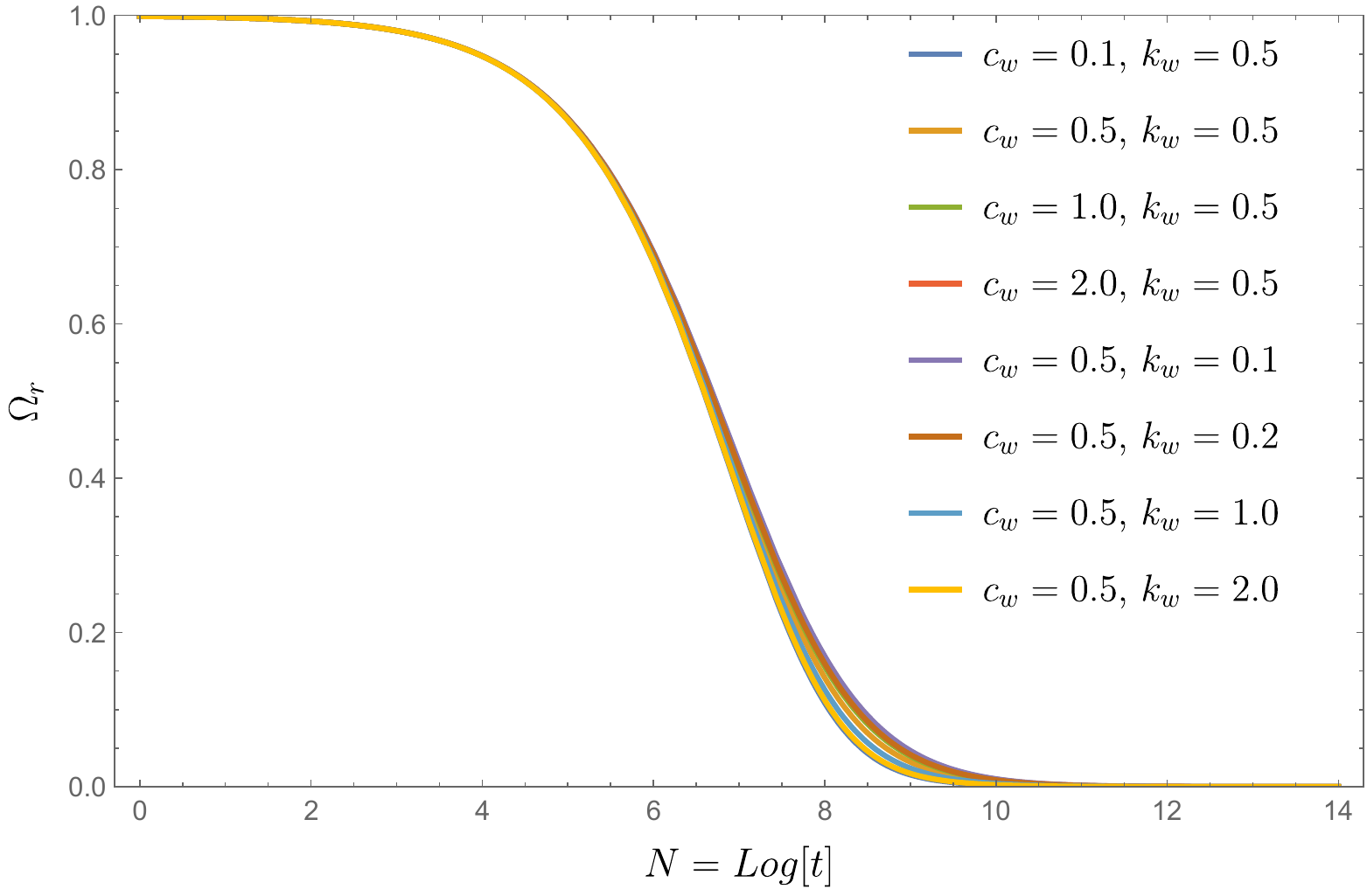}
    \caption{Parameter scan of $\Omega_w(N)$ for varying $c_w$ (fixed $k_w = 0.5$) and varying $k_w$ (fixed $c_w = 0.5$).}
    \label{fig:param_scan}
\end{figure}

To test robustness we vary the VOS parameters across $c_w \in \{0.1, 0.5, 1.0, 2.0\}$ (with $k_w = 0.5$) and $k_w \in \{0.1, 0.2, 1.0, 2.0\}$ (with $c_w = 0.5$). Fig.~\ref{fig:param_scan} shows that larger $c_w$ or smaller $k_w$ delays the onset of wall domination, but every trajectory eventually departs from the RD saddle and reaches the runaway branch. This confirms that the global structure---the RD saddle and the runaway toward kinematic frustration---is robust for all $c_w, k_w > 0$, in agreement with the analytic no-go result.

\section{Conclusions and Discussion}
\label{sec:conclusions}

We have shown that including gravitational backreaction qualitatively alters the phase-space structure of the VOS model for DW networks. The radiation-era scaling solution --- a stable attractor on a fixed background --- becomes a saddle once the expansion rate $q\equiv Ht$ is promoted to a dynamical variable. Within the physical phase space $0\le\Omega_r\le1$, no stable fixed point exists. Numerical integration confirms that trajectories are generically driven toward a runaway branch with $v\to0$, $L\propto a$, and $q\to2$, corresponding to kinematic frustration.

A natural concern is whether the constant-parameter approximation $c_w, k_w = \text{const}$ drives the conclusion. The proof is structured as a reductio ad absurdum: if a scaling attractor existed in the interior region $0<\Omega_r<1$, the phenomenological parameters --- whatever their microphysical dependence on $v$ or the background --- would approach constant values at that fixed point, at which point the Hurwitz analysis applies and shows that no physically admissible stable solution exists. The $\mu=+1$ eigenvalue at the RD boundary is parameter-independent similarly. The instability is therefore not an artifact of the constant-parameter approximation but a robust consequence of the autonomous structure, provided the single-scale description remains valid.

It is also worth to address that the VOS model's applicability in the low-velocity, self-gravitating regime has not been verified by field-theory simulations. The FRW metric is retained as an effective large-scale geometry; whether anisotropic wall stresses invalidate the isotropic coarse-graining when $\omega_w=-2/3$ requires a quantitative study. 

A potential caveat is that the energy lost by the wall network may be distributed among several channels, including particle production~\cite{Gelmini:1988sf,Gouttenoire:2025ofv,Hassan:2024bvb,kaloper2026electromagneticcouplingsdarkdomain,PhysRevD.40.1002,Hiramatsu:2012gg} and gravitational waves\cite{Kawasaki_2011,cyr2025nearpeakspectrumgravitationalwaves,Hiramatsu_2010,Kawasaki_2011}. Nevertheless, our conclusion does not depend on the detailed energy-transfer pathway. As wall domination suppresses the RMS velocity, all velocity-dependent loss mechanisms become less efficient, while thermal interactions are further weakened by the rapidly decreasing temperature. Therefore, these effects are unlikely to restore a scaling attractor.

Although we have focused on radiation because any cosmologically viable period of domain-wall domination is expected to occur before BBN, the qualitative mechanism is more general. Replacing radiation by matter or another background fluid changes the quantitative location of the fixed points, but the loss of the scaling attractor originates from the self-consistent coupling between the defect network and the background expansion rather than from the specific equation of state.

The established background dynamics motivates a systematic study of cosmological perturbations in the frustrated phase~\cite{PhysRevD.60.043505,faroughy2018cosmologicalperturbationsuniversedomain}(If $\sigma_w < 1 \mathrm{MeV}^3$, the domain-wall network does not enter a frustrated regime until now). This includes analyzing the evolution of superhorizon curvature perturbations and the subsequent reheating transition triggered by the decay of the frozen network.

\section*{Acknowledgments}
The author thanks Sabir Ramazanov for useful discussions and valuable comments on the manuscript. The author acknowledges the support from the Foundation for the Advancement of Theoretical Physics and Mathematics “BASIS”.

\appendix

\section{Derivation of the 4D Autonomous System}
\label{app:derivation}

Starting from the VOS equations~(\ref{eq:vos-L})--(\ref{eq:vos-v}) and the continuity equations~(\ref{eq:cont-w})--(\ref{eq:cont-r}), we change the time variable from cosmic time $t$ to $N = \ln t$. For any quantity $X$,
\begin{equation}
\frac{\d X}{\d t} = \frac{1}{t}\frac{\d X}{\d N},
\end{equation}
and since $H = q/t$, we have $\d/\d t = H q^{-1} \d/\d N$.

\subsection*{The $q$ equation}

From the Friedmann equation $H^2 = (8\pi G/3)(\rho_w + \rho_r)$ and the total continuity equation:
\begin{equation}
\dot\rho_{\rm tot} + 3H(\rho_{\rm tot} + p_{\rm tot}) = 0,
\end{equation}
with $p_{\rm tot} = \omega_w\rho_w + (1/3)\rho_r$ and $\omega_w = v^2 - 2/3$. Using $\rho_{\rm tot} = 3H^2/8\pi G$ and changing to $N$:
\begin{equation}
\frac{\d H}{\d N} = -\frac{3}{2}H\left[1 + \omega_w(1-\Omega_r) + \frac{1}{3}\Omega_r\right].
\end{equation}

Now $q = Ht = H e^N$. Differentiating:
\begin{equation}
\frac{\d q}{\d N} = \frac{\d H}{\d N}t + Ht =  q\left[1 - \frac{3}{2}\left(1 + \omega_w(1-\Omega_r) + \frac{1}{3}\Omega_r\right)\right].
\end{equation}

Substituting $\omega_w = v^2 - 2/3$ and simplifying yields the $q'$ equation in~(\ref{eq:4d-system}).

\subsection*{The $y$ equation}

From $y = HL$:
\begin{equation}
y' = \frac{\d y}{\d N} = \frac{\d H}{\d N}L + H\frac{\d L}{\d N} = q\frac{\d\ln H}{\d N}y + H\frac{\d L}{\d N}.
\end{equation}

Using $\d L/\d N = t \d L/\d t = t[HL(1+3v^2) + c_w v] = q y(1+3v^2) + c_w v t$. Since $t = e^N = q/H$, the $c_w v t$ term becomes $c_w v q/H$. This gives:
\begin{equation}
y' = -\frac{3}{2}q y\left[1 + \omega_w(1-\Omega_r) + \frac{1}{3}\Omega_r\right] + q\left[y(1+3v^2) + \frac{c_w v}{H}\right].
\end{equation}

Simplifying and substituting $\omega_w = v^2 - 2/3$ yields the $y'$ equation in~(\ref{eq:4d-system}).

\subsection*{The $\Omega_r$ equation}

From the continuity equations~(\ref{eq:cont-w})--(\ref{eq:cont-r}) and the definition $\Omega_r = \rho_r/(\rho_w + \rho_r)$:
\begin{align}
\Omega_r' &= 3q\Omega_r(1-\Omega_r)\left(\omega_w - \frac{1}{3}\right) \\
&+ 3q\Omega_r\left(\frac{1}{3} - \omega_w\right) + \frac{c_w v q}{y}(1-\Omega_r) \nonumber\\
&= q(1-\Omega_r)\Bigl[\frac{c_w v}{y} - 3\Omega_r(1-v^2)\Bigr],
\end{align}
after substituting $\omega_w = v^2 - 2/3$ and simplifying.

\section{Stability of the RD Fixed Point}
\label{app:rd-stability}

At $\Omega_r = 1$, the autonomous system~(\ref{eq:4d-system}) admits the RD scaling fixed point:
\begin{equation}
\Omega_{r*} = 1, \qquad
q_* = \frac{1}{2}, \qquad
v_* = \sqrt{\frac{k_w}{3(k_w + c_w)}}, \qquad
y_* = \frac{k_w}{3v_*}.
\label{eq:fp-rd-values}
\end{equation}
Evaluating the $4 \times 4$ Jacobian $M_{ij} = \partial X_i' / \partial X_j$ with $X = (y, v, q, \Omega_r)$ at this point and simplifying, the non-zero entries are:
\begin{align}
M_{11} &= -\frac{c_w}{2(c_w+k_w)}, \nonumber\\
M_{12} &= \frac{c_w}{2} + k_w, \nonumber\\
M_{13} &= -\frac{3c_w+2k_w}{4}\sqrt{\frac{k_w}{3(c_w+k_w)}}, \nonumber\\
M_{21} &= -\frac{3c_w+2k_w}{2(c_w+k_w)^2}, \nonumber\\
M_{22} &= -\frac{3c_w+2k_w}{2(c_w+k_w)}, \nonumber\\
M_{33} &= 1, \qquad M_{43} = -\frac{3c_w+2k_w}{8(c_w+k_w)}, \qquad M_{44} = -1,
\end{align}
with $M_{14}=M_{23}=M_{24}=M_{31}=M_{32}=M_{34}=M_{41}=M_{42}=0$.

The Jacobian is partially triangular. The lower-right $2 \times 2$ block yields eigenvalues $M_{33} = 1$ and $M_{44} = -1$ directly. The characteristic polynomial of the full matrix factorizes, giving the four eigenvalues analytically:
\begin{align}
\mu_1 &= -1, \nonumber\\
\mu_2 &= +1, \nonumber\\
\mu_{3,4} &= -1 + \frac{k_w}{2(c_w+k_w)}
\pm i\,\frac{\sqrt{2c_w^2 + 6c_w k_w + 3k_w^2}}{2(c_w+k_w)}.
\label{eq:ev-rd}
\end{align}

Since $c_w, k_w > 0$, the argument of the square root is strictly positive, so $\mu_{3,4}$ form a complex-conjugate pair. Their common real part satisfies $\Re(\mu_{3,4}) = -1 + k_w/[2(c_w + k_w)] < -1/2 < 0$. The spectrum is therefore $\{+1, -1, \lambda, \bar{\lambda}\}$ with $\Re(\lambda) < 0$ for all $c_w, k_w > 0$.

The fixed point is a saddle: it possesses one positive real eigenvalue ($\mu_2 = +1$) and three eigenvalues with negative real parts. Equivalently, the characteristic polynomial $\mu^4 + a_1\mu^3 + a_2\mu^2 + a_3\mu + a_4$ has coefficients $a_3 = -1 - c_w/(c_w+k_w) < 0$ and $a_4 = -1 - c_w/[2(c_w+k_w)] < 0$, immediately violating the Routh-Hurwitz conditions. Enforcing all six Hurwitz inequalities yields no solution for any $c_w, k_w > 0$.

The eigenvalue $\mu_2 = +1$ is independent of the model parameters and originates from $M_{33} = \partial q'/\partial q$ --- the self-coupling of the expansion rate variable. The upper-left $2 \times 2$ block (the $(y, v)$ subspace) reproduces the stable VOS attractor of the fixed-background analysis (Sec.~\ref{sec:vos-fixed}). Thus, on a fixed radiation background the RD point is a genuine attractor, but in the self-gravitating system the dynamical $q$ degree of freedom introduces an unstable direction, converting the attractor into a saddle. Trajectories near this point are attracted along the stable manifold but ultimately repelled, driving the system away from $\Omega_r = 1$ toward the runaway regime.

\section{Linear Stability Analysis of the Interior Fixed Point}
\label{app:jacobian-interior}

At the interior fixed point defined by Eqs.~(\ref{eq:omegar-interior})--(\ref{eq:q-interior}) with $y_* = k_w/(3v)$ and $q_* = 2k_w/[k_w + 3v^2(k_w + c_w)]$, we compute the Jacobian $J_{ij} = \partial X_i'/\partial X_j$ with $X = (y, v, q, \Omega_r)$. Defining $\Delta \equiv k_w + 3v^2(k_w + c_w)$, the non-zero entries are:
\begin{align}
J_{11} &= \frac{k_w(1+3v^2) - 3v^2 c_w}{\Delta}, \nonumber\\
J_{12} &= \frac{2k_w[k_w(1-v^2) + c_w]}{\Delta(1-v^2)}, \nonumber\\
J_{13} &= \frac{\Delta}{6v}, \nonumber\\
J_{14} &= -\frac{k_w^2(1-v^2)}{\Delta v}, \nonumber\\
J_{21} &= -\frac{18v^2(1-v^2)}{\Delta}, \nonumber\\
J_{22} &= -\frac{6k_w(1-v^2)}{\Delta}, \nonumber\\
J_{31} &= \frac{18v^3 c_w[v^2 c_w - (1-v^2)k_w]}{\Delta(1-v^2)k_w^2}, \nonumber\\
J_{32} &= -\frac{6v(1+v^2) c_w[v^2 c_w - (1-v^2)k_w]}{\Delta(1-v^2)^2 k_w}, \nonumber\\
J_{33} &= \frac{6[v^2 c_w - (1-v^2)k_w]}{\Delta}, \nonumber\\
J_{42} &= \frac{12v k_w[v^2 c_w - (1-v^2)k_w]}{\Delta^2(1-v^2)}, \nonumber\\
J_{43} &= -\frac{6k_w^2(1-v^2)}{\Delta^2}, \nonumber\\
J_{44} &= -1.
\end{align}
All other entries vanish: $J_{23}=J_{24}=J_{34}=J_{41}=0$.

The characteristic polynomial is $\det(J - \mu I) = \mu^4 + a_1 \mu^3 + a_2 \mu^2 + a_3 \mu + a_4$, with coefficients:
\begin{align}
a_1 &= \frac{12(1-v^2)k_w}{\Delta}, \nonumber\\[4pt]
a_2 &= \frac{-9v^4 c_w^2 + 18v^2(3-v^2)c_w k_w + (35-42v^2-9v^4)k_w^2}{\Delta^2}, \nonumber\\[4pt]
a_3 &= \frac{6}{\Delta^3}\Bigl[9v^6 c_w^3 + 12v^4(6v^2-1)c_w^2 k_w + v^2(37-90v^2+117v^4)c_w k_w^2 \nonumber\\
    &\qquad + 2(27v^6-39v^4+13v^2-1)k_w^3\Bigr], \\[4pt]
a_4 &= \frac{36(1-v^2)k_w\bigl[v^2 c_w - (1-v^2)k_w\bigr]}{\Delta^2},
\end{align}
where $\Delta \equiv k_w + 3v^2(k_w + c_w)$. The Hurwitz determinants are $H_1 \equiv a_1 a_2 - a_3$ and $H_2 \equiv a_1 a_2 a_3 - a_1^2 a_4 - a_3^2$.

A direct symbolic solution of all six Hurwitz inequalities with symbolic $c_w, k_w$ does not terminate. Proceeding hierarchically:

\begin{enumerate}
    \item $a_1, a_2, a_3, a_4 > 0$ with $0 < v < 1$, $c_w, k_w > 0$ yields $0 < v < \sqrt{2/3}$ with $k_w < v^2 c_w/(1-v^2)$.
    \item Adding $H_1 > 0$ constrains $k_w$ from below via a cubic Root object.
    \item Adding $H_2 > 0$ preserves the same structure.
\end{enumerate}

The necessary condition for stability is therefore
\begin{equation}
k_w < \frac{v^2 c_w}{1 - v^2}.
\label{eq:kw-upper-bound}
\end{equation}
Combining with the fixed-point relation $\Omega_r = c_w v^2 / [k_w(1-v^2)]$ yields $\Omega_r > 1$. Thus any mathematically stable interior fixed point lies strictly outside the physical phase space. 

\bibliographystyle{apsrev4-2}
\bibliography{references}

@article{Kibble:1976sj,
    author = "Kibble, T. W. B.",
    title = "{Topology of Cosmic Domains and Strings}",
    reportNumber = "ICTP/75/5",
    doi = "10.1088/0305-4470/9/8/029",
    journal = "J. Phys. A",
    volume = "9",
    pages = "1387--1398",
    year = "1976"
}

@article{Vilenkin:1984ib,
    author = "Vilenkin, Alexander",
    title = "{Cosmic Strings and Domain Walls}",
    reportNumber = "PRINT-84-0840 (TUFTS)",
    doi = "10.1016/0370-1573(85)90033-X",
    journal = "Phys. Rept.",
    volume = "121",
    pages = "263--315",
    year = "1985"
}

@article{PhysRevD.23.852,
  title = {Gravitational field of vacuum domain walls and strings},
  author = {Vilenkin, Alexander},
  journal = {Phys. Rev. D},
  volume = {23},
  issue = {4},
  pages = {852--857},
  numpages = {0},
  year = {1981},
  month = {Feb},
  publisher = {American Physical Society},
  doi = {10.1103/PhysRevD.23.852},
  url = {https://link.aps.org/doi/10.1103/PhysRevD.23.852}
}

@article{Hiramatsu_2013,
   title={Axion cosmology with long-lived domain walls},
   volume={2013},
   ISSN={1475-7516},
   url={http://dx.doi.org/10.1088/1475-7516/2013/01/001},
   DOI={10.1088/1475-7516/2013/01/001},
   number={01},
   journal={Journal of Cosmology and Astroparticle Physics},
   publisher={IOP Publishing},
   author={Hiramatsu, Takashi and Kawasaki, Masahiro and Saikawa, Ken’ichi and Sekiguchi, Toyokazu},
   year={2013},
   month=Jan, pages={001–001} }

@book{Vilenkin:2000jqa,
    author = "Vilenkin, A. and Shellard, E. P. S.",
    title = "{Cosmic Strings and Other Topological Defects}",
    isbn = "978-0-521-65476-0",
    publisher = "Cambridge University Press",
    month = "7",
    year = "2000"
}

@article{Linde:1990yj,
    author = "Linde, Andrei D. and Lyth, David H.",
    title = "{Axionic domain wall production during inflation}",
    reportNumber = "CERN-TH-5690-90",
    doi = "10.1016/0370-2693(90)90613-B",
    journal = "Phys. Lett. B",
    volume = "246",
    pages = "353--358",
    year = "1990"
}

@article{Hiramatsu_2011,
   title={Evolution of string-wall networks and axionic domain wall problem},
   volume={2011},
   ISSN={1475-7516},
   url={http://dx.doi.org/10.1088/1475-7516/2011/08/030},
   DOI={10.1088/1475-7516/2011/08/030},
   number={08},
   journal={Journal of Cosmology and Astroparticle Physics},
   publisher={IOP Publishing},
   author={Hiramatsu, Takashi and Kawasaki, Masahiro and Saikawa, Ken’ichi},
   year={2011},
   month=Aug, pages={030–030} }

@article{Sikivie:1982qv,
    author = "Sikivie, P.",
    title = "{Of Axions, Domain Walls and the Early Universe}",
    reportNumber = "UFTP-82-3",
    doi = "10.1103/PhysRevLett.48.1156",
    journal = "Phys. Rev. Lett.",
    volume = "48",
    pages = "1156--1159",
    year = "1982"
}

@article{Kawasaki:2014sqa,
    author = "Kawasaki, Masahiro and Saikawa, Ken'ichi and Sekiguchi, Toyokazu",
    title = "{Axion dark matter from topological defects}",
    eprint = "1412.0789",
    archivePrefix = "arXiv",
    primaryClass = "hep-ph",
    reportNumber = "ICRR-REPORT-696-2014-22, IPMU14-0348",
    doi = "10.1103/PhysRevD.91.065014",
    journal = "Phys. Rev. D",
    volume = "91",
    number = "6",
    pages = "065014",
    year = "2015"
}

@article{Hiramatsu:2012gg,
    author = "Hiramatsu, Takashi and Kawasaki, Masahiro and Saikawa, Ken'ichi and Sekiguchi, Toyokazu",
    title = "{Production of dark matter axions from collapse of string-wall systems}",
    eprint = "1202.5851",
    archivePrefix = "arXiv",
    primaryClass = "hep-ph",
    reportNumber = "ICRR-REPORT-608-2011-25, IPMU12-0025, YITP-12-9",
    doi = "10.1103/PhysRevD.85.105020",
    journal = "Phys. Rev. D",
    volume = "85",
    pages = "105020",
    year = "2012",
    note = "[Erratum: Phys.Rev.D 86, 089902 (2012)]"
}

@article{Bhattacharjee:1991zm,
    author = "Bhattacharjee, Pijushpani and Hill, Christopher T. and Schramm, David N.",
    title = "{Grand Unified Theories, Topological Defects and Ultrahigh-Energy Cosmic Rays}",
    reportNumber = "FERMILAB-PUB-91-304-A",
    doi = "10.1103/PhysRevLett.69.567",
    journal = "Phys. Rev. Lett.",
    volume = "69",
    pages = "567--570",
    year = "1992"
}

@article{Chakrabortty:2019fov,
    author = "Chakrabortty, Joydeep and Maji, Rinku and King, Stephen F.",
    title = "{Unification, Proton Decay and Topological Defects in non-SUSY GUTs with Thresholds}",
    eprint = "1901.05867",
    archivePrefix = "arXiv",
    primaryClass = "hep-ph",
    doi = "10.1103/PhysRevD.99.095008",
    journal = "Phys. Rev. D",
    volume = "99",
    number = "9",
    pages = "095008",
    year = "2019"
}

@article{Bajc:1997ky,
    author = "Bajc, Borut and Riotto, Antonio and Senjanovic, Goran",
    title = "{Large lepton number of the universe and the fate of topological defects}",
    eprint = "hep-ph/9710415",
    archivePrefix = "arXiv",
    reportNumber = "DTP-97-92, IJS-TP-97-16, OUTP-97-49-P",
    doi = "10.1103/PhysRevLett.81.1355",
    journal = "Phys. Rev. Lett.",
    volume = "81",
    pages = "1355--1358",
    year = "1998"
}

@article{King_2021,
   title={Gravitational Waves and Proton Decay: Complementary Windows into Grand Unified Theories},
   volume={126},
   ISSN={1079-7114},
   url={http://dx.doi.org/10.1103/PhysRevLett.126.021802},
   DOI={10.1103/physrevlett.126.021802},
   number={2},
   journal={Physical Review Letters},
   publisher={American Physical Society (APS)},
   author={King, Stephen F. and Pascoli, Silvia and Turner, Jessica and Zhou, Ye-Ling},
   year={2021},
   month=Jan }

@article{Abel:1995wk,
    author = "Abel, S. A. and Sarkar, Subir and White, P. L.",
    title = "{On the cosmological domain wall problem for the minimally extended supersymmetric standard model}",
    eprint = "hep-ph/9506359",
    archivePrefix = "arXiv",
    reportNumber = "OUTP-95-22-P, RAL-TR-95-019",
    doi = "10.1016/0550-3213(95)00483-9",
    journal = "Nucl. Phys. B",
    volume = "454",
    pages = "663--684",
    year = "1995"
}

@article{Preskill:1991kd,
    author = "Preskill, John and Trivedi, Sandip P. and Wilczek, Frank and Wise, Mark B.",
    title = "{Cosmology and broken discrete symmetry}",
    reportNumber = "IASSNS-HEP-91-11, CALT-68-1718",
    doi = "10.1016/0550-3213(91)90241-O",
    journal = "Nucl. Phys. B",
    volume = "363",
    pages = "207--220",
    year = "1991"
}

@article{Gelmini:1988sf,
    author = "Gelmini, Graciela B. and Gleiser, Marcelo and Kolb, Edward W.",
    title = "{Cosmology of Biased Discrete Symmetry Breaking}",
    reportNumber = "NSF-ITP-88-148, FERMILAB-PUB-88-151-A",
    doi = "10.1103/PhysRevD.39.1558",
    journal = "Phys. Rev. D",
    volume = "39",
    pages = "1558",
    year = "1989"
}

@article{Battye:2011jj,
    author = "Battye, Richard A. and Brawn, Gary D. and Pilaftsis, Apostolos",
    title = "{Vacuum Topology of the Two Higgs Doublet Model}",
    eprint = "1106.3482",
    archivePrefix = "arXiv",
    primaryClass = "hep-ph",
    doi = "10.1007/JHEP08(2011)020",
    journal = "JHEP",
    volume = "08",
    pages = "020",
    year = "2011"
}

@article{PhysRevD.28.1419,
  title = {Domain walls. II. Baryon-number generation},
  author = {Holdom, Bob},
  journal = {Phys. Rev. D},
  volume = {28},
  issue = {6},
  pages = {1419--1424},
  numpages = {0},
  year = {1983},
  month = {Sep},
  publisher = {American Physical Society},
  doi = {10.1103/PhysRevD.28.1419},
  url = {https://link.aps.org/doi/10.1103/PhysRevD.28.1419}
}

@article{Lalak:1993ay,
    author = "Lalak, Zygmunt and Thomas, Steven",
    title = "{Domain wall formation in the postinflationary universe}",
    eprint = "hep-ph/9303250",
    archivePrefix = "arXiv",
    reportNumber = "QMW-PH-93-4",
    doi = "10.1016/0370-2693(93)91130-F",
    journal = "Phys. Lett. B",
    volume = "306",
    pages = "10--18",
    year = "1993"
}

@article{Gonzalez:2022mcx,
    author = "Gonzalez, Diego and Kitajima, Naoya and Takahashi, Fuminobu and Yin, Wen",
    title = "{Stability of domain wall network with initial inflationary fluctuations and its implications for cosmic birefringence}",
    eprint = "2211.06849",
    archivePrefix = "arXiv",
    primaryClass = "hep-ph",
    reportNumber = "TU-1174",
    doi = "10.1016/j.physletb.2023.137990",
    journal = "Phys. Lett. B",
    volume = "843",
    pages = "137990",
    year = "2023"
}

@article{Babichev:2021uvl,
    author = "Babichev, E. and Gorbunov, D. and Ramazanov, S. and Vikman, A.",
    title = "{Gravitational shine of dark domain walls}",
    eprint = "2112.12608",
    archivePrefix = "arXiv",
    primaryClass = "hep-ph",
    doi = "10.1088/1475-7516/2022/04/028",
    journal = "JCAP",
    volume = "04",
    number = "04",
    pages = "028",
    year = "2022"
}

@article{Babichev:2025stm,
    author = "Babichev, E. and Dankovsky, I. and Gorbunov, D. and Ramazanov, S. and Vikman, A.",
    title = "{Biased domain walls: faster annihilation, weaker gravitational waves}",
    eprint = "2504.07902",
    archivePrefix = "arXiv",
    primaryClass = "hep-ph",
    doi = "10.1088/1475-7516/2025/10/103",
    journal = "JCAP",
    volume = "10",
    pages = "103",
    year = "2025"
}

@article{Wei:2022poh,
    author = "Wei, Dongdong and Jiang, Yun",
    title = "{Domain wall networks from first-order phase transitions and gravitational waves}",
    eprint = "2208.07186",
    archivePrefix = "arXiv",
    primaryClass = "hep-ph",
    doi = "10.1103/PhysRevD.110.123505",
    journal = "Phys. Rev. D",
    volume = "110",
    number = "12",
    pages = "123505",
    year = "2024"
}

@article{Blasi_2022,
   title={Domain Walls Seeding the Electroweak Phase Transition},
   volume={129},
   ISSN={1079-7114},
   url={http://dx.doi.org/10.1103/PhysRevLett.129.261303},
   DOI={10.1103/physrevlett.129.261303},
   number={26},
   journal={Physical Review Letters},
   publisher={American Physical Society (APS)},
   author={Blasi, Simone and Mariotti, Alberto},
   year={2022},
   month=Dec }

@article{Li:2023tdx,
    author = "Li, Xiu-Fei",
    title = "{Probing the high temperature symmetry breaking with gravitational waves from domain walls}",
    eprint = "2307.03163",
    archivePrefix = "arXiv",
    primaryClass = "hep-ph",
    doi = "10.1016/j.nuclphysb.2025.117036",
    journal = "Nucl. Phys. B",
    volume = "1018",
    pages = "117036",
    year = "2025"
}

@article{Zeldovich:1974uw,
    author = "Zeldovich, Ya. B. and Kobzarev, I. Yu. and Okun, L. B.",
    title = "{Cosmological Consequences of the Spontaneous Breakdown of Discrete Symmetry}",
    reportNumber = "SLAC-TRANS-0165, IPM-MOSCOW-15",
    journal = "Zh. Eksp. Teor. Fiz.",
    volume = "67",
    pages = "3--11",
    year = "1974"
}

@article{Avelino_2005,
   title={One-scale model for domain wall network evolution},
   volume={72},
   ISSN={1550-2368},
   url={http://dx.doi.org/10.1103/PhysRevD.72.083506},
   DOI={10.1103/physrevd.72.083506},
   number={8},
   journal={Physical Review D},
   publisher={American Physical Society (APS)},
   author={Avelino, P. P. and Martins, C. J. A. P. and Oliveira, J. C. R. E.},
   year={2005},
   month=Oct }

@article{Avelino_2006,
   title={Defect junctions and domain wall dynamics},
   volume={73},
   ISSN={1550-2368},
   url={http://dx.doi.org/10.1103/PhysRevD.73.123520},
   DOI={10.1103/physrevd.73.123520},
   number={12},
   journal={Physical Review D},
   publisher={American Physical Society (APS)},
   author={Avelino, P. P. and Martins, C. J. A. P. and Menezes, J. and Menezes, R. and Oliveira, J. C. R. E.},
   year={2006},
   month=June }

@article{PhysRevD.53.4237,
  title = {Biased domain walls},
  author = {Coulson, D. and Lalak, Z. and Ovrut, B.},
  journal = {Phys. Rev. D},
  volume = {53},
  issue = {8},
  pages = {4237--4246},
  numpages = {0},
  year = {1996},
  month = {Apr},
  publisher = {American Physical Society},
  doi = {10.1103/PhysRevD.53.4237},
  url = {https://link.aps.org/doi/10.1103/PhysRevD.53.4237}
}

@article{Kuroyanagi_2018,
   title={Probing the Universe through the stochastic gravitational wave background},
   volume={2018},
   ISSN={1475-7516},
   url={http://dx.doi.org/10.1088/1475-7516/2018/11/038},
   DOI={10.1088/1475-7516/2018/11/038},
   number={11},
   journal={Journal of Cosmology and Astroparticle Physics},
   publisher={IOP Publishing},
   author={Kuroyanagi, Sachiko and Chiba, Takeshi and Takahashi, Tomo},
   year={2018},
   month=Nov, pages={038–038} }

@article{Saikawa_2017,
   title={A Review of Gravitational Waves from Cosmic Domain Walls},
   volume={3},
   ISSN={2218-1997},
   url={http://dx.doi.org/10.3390/universe3020040},
   DOI={10.3390/universe3020040},
   number={2},
   journal={Universe},
   publisher={MDPI AG},
   author={Saikawa, Ken’ichi},
   year={2017},
   month=May, pages={40} }

@article{Dankovsky:2024ipq,
    author = "Dankovsky, I. and Ramazanov, S. and Babichev, E. and Gorbunov, D. and Vikman, A.",
    title = "{Numerical analysis of melting domain walls and their gravitational waves}",
    eprint = "2410.21971",
    archivePrefix = "arXiv",
    primaryClass = "hep-ph",
    doi = "10.1088/1475-7516/2025/02/064",
    journal = "JCAP",
    volume = "02",
    pages = "064",
    year = "2025"
}

@article{Dankovsky:2024zvs,
    author = "Dankovsky, I. and Babichev, E. and Gorbunov, D. and Ramazanov, S. and Vikman, A.",
    title = "{Revisiting evolution of domain walls and their gravitational radiation with CosmoLattice}",
    eprint = "2406.17053",
    archivePrefix = "arXiv",
    primaryClass = "astro-ph.CO",
    doi = "10.1088/1475-7516/2024/09/047",
    journal = "JCAP",
    volume = "09",
    pages = "047",
    year = "2024"
}

@article{Chang_1998,
   title={Studies of the motion and decay of axion walls bounded by strings},
   volume={59},
   ISSN={1089-4918},
   url={http://dx.doi.org/10.1103/PhysRevD.59.023505},
   DOI={10.1103/physrevd.59.023505},
   number={2},
   journal={Physical Review D},
   publisher={American Physical Society (APS)},
   author={Chang, S. and Hagmann, C. and Sikivie, P.},
   year={1998},
   month=Dec }

@article{Gleiser_1998,
   title={Gravitational Waves from Collapsing Vacuum Domains},
   volume={81},
   ISSN={1079-7114},
   url={http://dx.doi.org/10.1103/PhysRevLett.81.5497},
   DOI={10.1103/physrevlett.81.5497},
   number={25},
   journal={Physical Review Letters},
   publisher={American Physical Society (APS)},
   author={Gleiser, Marcelo and Roberts, Ronald},
   year={1998},
   month=Dec, pages={5497–5500} }

@article{Hiramatsu_2010,
   title={Gravitational waves from collapsing domain walls},
   volume={2010},
   ISSN={1475-7516},
   url={http://dx.doi.org/10.1088/1475-7516/2010/05/032},
   DOI={10.1088/1475-7516/2010/05/032},
   number={05},
   journal={Journal of Cosmology and Astroparticle Physics},
   publisher={IOP Publishing},
   author={Hiramatsu, Takashi and Kawasaki, Masahiro and Saikawa, Ken’ichi},
   year={2010},
   month=May, pages={032–032} }

@misc{roshan2026imprintdomainwallannihilation,
      title={Imprint of domain wall annihilation on induced gravitational waves}, 
      author={Rishav Roshan},
      year={2026},
      eprint={2604.25726},
      archivePrefix={arXiv},
      primaryClass={hep-ph},
      url={https://arxiv.org/abs/2604.25726}, 
}

@misc{blasi2025domainwallsscalingregime,
      title={Domain walls in the scaling regime: Equal Time Correlator and Gravitational Waves}, 
      author={Simone Blasi and Alberto Mariotti and Aäron Rase and Miguel Vanvlasselaer},
      year={2025},
      eprint={2511.16649},
      archivePrefix={arXiv},
      primaryClass={hep-ph},
      url={https://arxiv.org/abs/2511.16649}, 
}

@article{Notari:2025kqq,
    author = "Notari, Alessio and Rompineve, Fabrizio and Torrenti, Francisco",
    title = "{The spectrum of gravitational waves from annihilating domain walls}",
    eprint = "2504.03636",
    archivePrefix = "arXiv",
    primaryClass = "astro-ph.CO",
    doi = "10.1088/1475-7516/2025/07/049",
    journal = "JCAP",
    volume = "07",
    pages = "049",
    year = "2025"
}

@ARTICLE{1989ApJ...347..590P,
       author = {{Press}, William H. and {Ryden}, Barbara S. and {Spergel}, David N.},
        title = "{Dynamical Evolution of Domain Walls in an Expanding Universe}",
      journal = {\apj},
         year = 1989,
        month = dec,
       volume = {347},
        pages = {590},
          doi = {10.1086/168151},
       adsurl = {https://ui.adsabs.harvard.edu/abs/1989ApJ...347..590P}
}

@article{PhysRevD.84.103523,
  title = {Scaling properties of domain wall networks},
  author = {Leite, A. M. M. and Martins, C. J. A. P.},
  journal = {Phys. Rev. D},
  volume = {84},
  issue = {10},
  pages = {103523},
  numpages = {9},
  year = {2011},
  month = {Nov},
  publisher = {American Physical Society},
  doi = {10.1103/PhysRevD.84.103523},
  url = {https://link.aps.org/doi/10.1103/PhysRevD.84.103523}
}

@article{Garagounis_2003,
   title={Scaling in numerical simulations of domain walls},
   volume={68},
   ISSN={1089-4918},
   url={http://dx.doi.org/10.1103/PhysRevD.68.103506},
   DOI={10.1103/physrevd.68.103506},
   number={10},
   journal={Physical Review D},
   publisher={American Physical Society (APS)},
   author={Garagounis, Theodore and Hindmarsh, Mark},
   year={2003},
   month=Nov }

@article{Oliveira_2005,
   title={Cosmological evolution of domain wall networks},
   volume={71},
   ISSN={1550-2368},
   url={http://dx.doi.org/10.1103/PhysRevD.71.083509},
   DOI={10.1103/physrevd.71.083509},
   number={8},
   journal={Physical Review D},
   publisher={American Physical Society (APS)},
   author={Oliveira, J. C. R. E. and Martins, C. J. A. P. and Avelino, P. P.},
   year={2005},
   month=Apr }

@article{Leite_2013,
   title={Accurate calibration of the velocity-dependent one-scale model for domain walls},
   volume={718},
   ISSN={0370-2693},
   url={http://dx.doi.org/10.1016/j.physletb.2012.11.070},
   DOI={10.1016/j.physletb.2012.11.070},
   number={3},
   journal={Physics Letters B},
   publisher={Elsevier BV},
   author={Leite, A.M.M. and Martins, C.J.A.P. and Shellard, E.P.S.},
   year={2013},
   month=Jan, pages={740–744} }

@article{Martins_2016,
   title={Extending the velocity-dependent one-scale model for domain walls},
   volume={93},
   ISSN={2470-0029},
   url={http://dx.doi.org/10.1103/PhysRevD.93.043534},
   DOI={10.1103/physrevd.93.043534},
   number={4},
   journal={Physical Review D},
   publisher={American Physical Society (APS)},
   author={Martins, C. J. A. P. and Rybak, I. Yu. and Avgoustidis, A. and Shellard, E. P. S.},
   year={2016},
   month=Feb }

@misc{avelino2022analyticalscalingsolutionsevolution,
      title={Analytical scaling solutions for the evolution of cosmic domain walls in a parameter-free velocity-dependent one-scale model}, 
      author={P. P. Avelino and D. Grüber and L. Sousa},
      year={2022},
      eprint={2203.16173},
      archivePrefix={arXiv},
      primaryClass={astro-ph.CO},
      url={https://arxiv.org/abs/2203.16173}, 
}

@article{Hassan:2024bvb,
    author = "Hassan, Saquib and Kane, Gaurang Ramakant and March-Russell, John and Obied, Georges",
    title = "{Chern-Simons induced thermal friction on axion domain walls}",
    eprint = "2410.19906",
    archivePrefix = "arXiv",
    primaryClass = "hep-ph",
    doi = "10.1007/JHEP03(2025)022",
    journal = "JHEP",
    volume = "03",
    pages = "022",
    year = "2025"
}

@misc{kaloper2026electromagneticcouplingsdarkdomain,
      title={Electromagnetic Couplings of Dark Domain Walls}, 
      author={Nemanja Kaloper},
      year={2026},
      eprint={2602.03933},
      archivePrefix={arXiv},
      primaryClass={hep-th},
      url={https://arxiv.org/abs/2602.03933}, 
}

@article{Gouttenoire:2025ofv,
    author = "Gouttenoire, Yann and King, Stephen F. and Roshan, Rishav and Wang, Xin and White, Graham and Yamazaki, Masahito",
    title = "{Cosmological consequences of domain walls biased by quantum gravity}",
    eprint = "2501.16414",
    archivePrefix = "arXiv",
    primaryClass = "hep-ph",
    doi = "10.1103/7zmx-v16z",
    journal = "Phys. Rev. D",
    volume = "112",
    number = "7",
    pages = "075007",
    year = "2025"
}

@article{PhysRevD.40.1002,
  title = {Dynamics of thick domain walls},
  author = {Widrow, Lawrence M.},
  journal = {Phys. Rev. D},
  volume = {40},
  issue = {4},
  pages = {1002--1010},
  numpages = {0},
  year = {1989},
  month = {Aug},
  publisher = {American Physical Society},
  doi = {10.1103/PhysRevD.40.1002},
  url = {https://link.aps.org/doi/10.1103/PhysRevD.40.1002}
}

@article{Kawasaki_2011,
   title={Study of gravitational radiation from cosmic domain walls},
   volume={2011},
   ISSN={1475-7516},
   url={http://dx.doi.org/10.1088/1475-7516/2011/09/008},
   DOI={10.1088/1475-7516/2011/09/008},
   number={09},
   journal={Journal of Cosmology and Astroparticle Physics},
   publisher={IOP Publishing},
   author={Kawasaki, Masahiro and Saikawa, Ken’ichi},
   year={2011},
   month=Sept, pages={008–008} }

@article{PhysRevD.60.043505,
  title = {Is the dark matter a solid?},
  author = {Bucher, Martin and Spergel, David},
  journal = {Phys. Rev. D},
  volume = {60},
  issue = {4},
  pages = {043505},
  numpages = {11},
  year = {1999},
  month = {Jul},
  publisher = {American Physical Society},
  doi = {10.1103/PhysRevD.60.043505},
  url = {https://link.aps.org/doi/10.1103/PhysRevD.60.043505}
}

@misc{faroughy2018cosmologicalperturbationsuniversedomain,
      title={Cosmological Perturbations in a Universe with a Domain Wall Era}, 
      author={Cyrus Faroughy},
      year={2018},
      eprint={1812.02344},
      archivePrefix={arXiv},
      primaryClass={hep-ph},
      url={https://arxiv.org/abs/1812.02344}, 
}

@misc{ferreira2025axionicdefectscmbbirefringence,
      title={Axionic defects in the CMB: birefringence and gravitational waves}, 
      author={Ricardo Z. Ferreira and Silvia Gasparotto and Takashi Hiramatsu and Ippei Obata and Oriol Pujolas},
      year={2025},
      eprint={2312.14104},
      archivePrefix={arXiv},
      primaryClass={hep-ph},
      url={https://arxiv.org/abs/2312.14104}, 
}

@misc{cyr2025nearpeakspectrumgravitationalwaves,
      title={Near-Peak Spectrum of Gravitational Waves from Collapsing Domain Walls}, 
      author={Bryce Cyr and Steven Cotterill and Richard Battye},
      year={2025},
      eprint={2504.02076},
      archivePrefix={arXiv},
      primaryClass={astro-ph.CO},
      url={https://arxiv.org/abs/2504.02076}, 
}

@article{Lew_1993,
   title={Discrete quark-lepton symmetry need not pose a cosmological domain wall problem},
   volume={47},
   ISSN={0556-2821},
   url={http://dx.doi.org/10.1103/PhysRevD.47.1356},
   DOI={10.1103/physrevd.47.1356},
   number={4},
   journal={Physical Review D},
   publisher={American Physical Society (APS)},
   author={Lew, H. and Volkas, R. R.},
   year={1993},
   month=Feb, pages={1356–1369} }

@article{Dvali_1995,
   title={Is There a Domain Wall Problem?},
   volume={74},
   ISSN={1079-7114},
   url={http://dx.doi.org/10.1103/PhysRevLett.74.5178},
   DOI={10.1103/physrevlett.74.5178},
   number={26},
   journal={Physical Review Letters},
   publisher={American Physical Society (APS)},
   author={Dvali, Gia and Senjanović, Goran},
   year={1995},
   month=June, pages={5178–5181} }

\end{document}